\documentclass[prd,aps,twocolumn,showpacs,preprintnumbers,amsmath,amssymb,superscriptaddress,nofootinbib,english]{revtex4-1}
\usepackage[dvips]{graphicx}
\usepackage{epsf}
\usepackage{amsmath}
\usepackage{amssymb}
\usepackage{amsfonts}
\usepackage{epstopdf}
\usepackage{times}
\usepackage[usenames]{color}
\usepackage[normalem]{ulem}
\usepackage[T1]{fontenc}
\usepackage[dvipsnames]{xcolor}

\usepackage{graphicx}% Include figure files
\usepackage{dcolumn}% Align table columns on decimal point
\usepackage{bm}% bold math
\usepackage{color}

\begin{document}

\title{Effective dark matter power spectra in $f(R)$ gravity}

\author{Jian-hua He}
\email[Email address: ]{jianhua.he@brera.inaf.it}
\affiliation{INAF-Observatorio Astronomico, di Brera, Via Emilio Bianchi, 46, I-23807, Merate (LC), Italy}
\affiliation{Institute for Computational Cosmology, Department of Physics, Durham University, Durham DH1 3LE, UK}

\author{Baojiu Li}
\affiliation{Institute for Computational Cosmology, Department of Physics, Durham University, Durham DH1 3LE, UK}

\author{Adam~J.~Hawken}
\affiliation{INAF-Observatorio Astronomico, di Brera, Via Emilio Bianchi, 46, I-23807, Merate (LC), Italy}

%\author{.....}
%\affiliation{XXXXXXXXXX}


\begin{abstract}
Using N-body simulations, we measure the power spectrum of the effective dark matter density field, which is defined through the modified Poisson equation in $f(R)$ cosmologies. We find that when compared to the conventional dark matter power spectrum, the effective power spectrum deviates more significantly from the $\Lambda$CDM model. For models with $f_{R0}=-10^{-4}$, the deviation can exceed 150\% while the deviation of the conventional matter power spectrum is less than 50\%. Even for models with $f_{R0}=-10^{-6}$, for which the conventional matter power spectrum is very close to the $\Lambda$CDM prediction, the effective power spectrum shows sizeable deviations. Our results indicate that traditional analyses based on the dark matter density field may seriously underestimate the impact of $f(R)$ gravity on galaxy clustering. We therefore suggest the use of the effective density field in such studies. In addition, based on our findings, we also discuss several possible methods of making use of the differences between the conventional and effective dark matter power spectra in $f(R)$ gravity to discriminate the theory from the $\Lambda$CDM model.
\end{abstract}

\maketitle

\section{Introduction}
It is well established that the Universe is currently undergoing a period of accelerated expansion~\cite{1}. The standard paradigm holds that this accelerated expansion is caused by a non-zero cosmological constant. A much discussed alternative is that a modification to general relativity could also account for the observed acceleration. However, the dynamics of the solar system heavily constrain the form that any such modification may take. Any viable theory of gravity must be indistinguishable from standard general relativity in high density environments, such as the Solar System and the centers of galaxies. A viable theory must also reproduce the observed expansion history, which is very well fit by the standard $\Lambda$CDM cosmology~\cite{WMAP,planck1,BAOm}.

Chameleon $f(R)$ gravity is a class of models where the effective potential of the scalar field is dependent upon the local environment \cite{Khoury}. A fifth force is involved which can be "screened" by the local density field in very deep potential wells. In regions where the matter density is high and the potential well is very deep, the fifth force is usually screened and gravity behaves just like standard GR. However, in low density regions of space there is no such screening and the strength of gravity is enhanced by the fifth force.

There are several observational effects due to such a modification. However, in order to make competitive forecasts for constraining $f(R)$ gravity with current and future cosmological surveys, it will be necessary to study the clustering of galaxies and to produce mock galaxy catalogues from simulations in $f(R)$ gravity.

However, the clustering of galaxies in $f(R)$ gravity is very complicated. The gravitational force produced by the dark matter field is mediated by the fifth force, which is no longer the same as that in standard gravity. The relationship between dark matter halos and the clustering of galaxies is not the same in $f(R)$ gravity compared to the standard model. Although the clustering of dark matter halos in $f(R)$ gravity has already been studied in the literature~\cite{frclustering}, it might be a risk to use the standard dark matter halos in semi-analytical galaxy formation models~\cite{semi}
to analyze the formation and clustering of galaxies. However, if we define an effective density field, the gravitational force produced by the effective density field could still have the same form as that in standard gravity. The clustering of galaxies in the effective dark matter density field could still follow that in the standard gravity. It is therefore more interesting to study the effective density field than the standard dark matter density field when analyzing the formation and clustering of galaxies.

In Ref.~\cite{effhalos} using the effective density field, we studied the properties of effective halos. We found that the relationships between the effective mass of a halo and its dynamical properties closely resemble those in the $\Lambda$CDM cosmology. This is an interesting result. The aim of this paper is to further extend this idea. We shall not only focus on halos but also on the entire density field. We shall study the power spectrum of the effective density field since in $f(R)$ gravity it is closely related to the galaxy power spectrum which is one of the most important statistical measures of the clustering of galaxies in cosmology.

This paper is organized as follows: In Sec.~\ref{model}, we introduce the $f(R)$ model studied in this work. In Sec.~\ref{nonlinear}, we discuss the linear and non-linear perturbation equations in $f(R)$ gravity. We also present the technical details of our simulations. In Sec.~\ref{power}, we describe our method of measuring the power spectra of scalar fields and we also present the numerical results on the effective power spectra. In Sec.~\ref{conclusions}, we summarize and conclude this work.

\section{f(R) model\label{model}}

The $f(R)$ gravity model is defined with the four-dimensional modified Einstein-Hilbert action
\begin{equation}
S=\frac{1}{2\kappa^2}\int d^4x\sqrt{-g}[R+f(R)]+\int
d^4x\mathcal{L}^{(m)}\label{action}\quad,
\end{equation}
where $\kappa^2=8\pi G$ with $G$ being Newton's constant, $g$ is the determinant of the metric $g_{\mu\nu}$, $\mathcal{L}^{(m)}$ is the Lagrangian density for matter, and $f(R)$ is an arbitrary algebraic function of the Ricci scalar curvature $R$ \cite{fr1,fr2,fr3,fr4,fr5,fr6,fr7,fr8,fr9,fr10,fr11,fr12} (see Refs.~\cite{frreview,review_Tsujikawa} for reviews).

In this work, we consider an $f(R)$ model that can exactly reproduce the $\Lambda$CDM background expansion history~\cite{frmodel}.
\begin{equation}
\begin{split}
f(R)&=-6\Omega_d^0H_0^2-\frac{3D\Omega_m^0H_0^2}{p_+-1}\left (\frac{3\Omega_m^0H_0^2}{R-12\Omega_d^0H_0^2}\right )^{p_+-1}\\
&\times{_2F_1}\left[q_+,p_+-1;r_+;-\frac{3\Omega_d^0H_0^2}{R-12\Omega_d^0H_0^2}\right
]\label{fr}\quad.
\end{split}
\end{equation}
The indices in the above expression are given by
\begin{eqnarray}
q_+=\frac{1+\sqrt{73}}{12}\nonumber,\quad r_+=1+\frac{\sqrt{73}}{6}\nonumber, \quad p_+=\frac{5+\sqrt{73}}{12}\nonumber.
\end{eqnarray}
${_2F_1}\left[a,b;c;z\right]$ is the hypergeometric function.
When $c>b>0$, the hypergeometric function has the integral representation
\begin{equation}
{_2F_1}[a,b;c;z]=\frac{\Gamma(c)}{\Gamma(b)\Gamma(c-b)}\int_0^{1}t^{b-1}(1-t)^{c-b-1}(1-zt)^{-a}dt\quad,\label{defhypergeometric}
\end{equation}
where $\Gamma(x)$ is the Euler gamma function. ${_2F_1}[a,b;c;z]$ is a real function that is well defined in the range $-\infty<z<1$. $H_0$ is the present Hubble constant. $\Omega_m^0$ is the matter density today and $\Omega_d^0=1-\Omega_m^0$. $D$ is an additional parameter that characterises the $f(R)$ model. For the instability issue as discussed in Ref.~\cite{Ignacy}, $D$ should be constrained as $D<0$. The model predicts a lower bound for the scalar curvature $R$ across the Universe
\begin{equation}
R\in(4\Lambda,+\infty)\quad,
\end{equation}
where
\begin{equation}
\Lambda=3\Omega_d^0H_0^2\quad.
\end{equation}
\section{N-body simulations\label{nonlinear}}

In this section, we will briefly summarise the basic equations that are used in $f(R)$ cosmological simulations.

\subsection{Non-linear perturbation equations and screening mechanism}
The formation of large-scale structure in $f(R)$ gravity is governed by the modified Poisson equation
\begin{equation}
\nabla^2\phi=\frac{16\pi G}{3}\delta \rho-\frac{\delta R}{6}\quad,\label{poissonfr}
\end{equation}
as well as an equation for the scalar field $f_{R}\equiv\frac{df}{dR}$. If the amplitude of the scalar field $|f_{R}|$ is very small ($|f_{R}|\ll1$),  the equation of motion for $f_{R}$ can be presented as
\begin{equation}
\nabla^2\delta f_R=\frac{1}{3c^2}[\delta R - 8\pi G\delta \rho]\quad,\label{frpoisson}
\end{equation}
where $\phi$ denotes the gravitational potential, $\delta f_R=f_R(R)-f_R(\bar{R})$, $\delta R= R-\bar{R}$, and $\delta \rho=\rho-\bar{\rho}$. The overbar denotes the background quantities, and $\nabla$ is the derivative with respect to the physical coordinates. Combining Eq.~(\ref{poissonfr}) and Eq.~(\ref{frpoisson}), we have
\begin{equation}
\nabla^2\phi_N=4\pi G\delta \rho \label{poissonN}\quad,
\end{equation}
where
\begin{equation}
\phi_N=\phi+\frac{c^2\delta f_R}{2}\quad,\label{N_phi}
\end{equation}
is the lensing potential~\cite{HuS}. If we define an effective density field,  the modified Poisson equation, Eq.~(\ref{poissonfr}), can be cast into the same form as Eq.~(\ref{poissonN})
\begin{equation}
\nabla^2\phi=4\pi G \delta \rho_{\rm eff}\quad,\label{eff_poisson}
\end{equation}
where $\delta \rho_{\rm eff}\equiv \frac{G_{\rm eff}}{G}\delta \rho$ and  $G_{\rm eff}$ is the effective Newtonian constant which is defined by
\begin{equation}
G_{\rm eff}=\left(\frac{4}{3}-\frac{\delta R}{24\pi G\delta \rho}\right)G\, .
\end{equation}
$G_{\rm eff}$ characterizes the strength of gravity among massive particles in $f(R)$ gravity.

As is well known, in linear theory $f(R)$ gravity can be ruled out due to the enhancement of gravity relative to the standard gravity $G_{\rm eff}\sim \frac{4}{3}G$. This conclusion is  regardless of the functional form of $f(R)$ (also see the appendix~\ref{appA}). Thus, a screening mechanism is essential for $f(R)$ theories to evade stringent local tests of gravity. However, there are two aspects of the screening mechanism in $f(R)$ theory to consider:

\begin{itemize}
\item A high curvature, $R\sim 8\pi G \rho$, should be recovered in high-density regions.
\item The fifth force, $\nabla f_R$, should be sufficiently suppressed in high-density regions as well.
\end{itemize}
Given the fact that the functional form of $f(R)$ is usually chosen as $\lim_{R\rightarrow+\infty}|f_R|=0$, if a high curvature can be achieved in high-density regions, the fifth force will be automatically suppressed:$\nabla \delta f_R=\nabla f_R\rightarrow 0$. However, it is important to note that a high-density does not imply a high curvature in $f(R)$ gravity.  As is discussed in Ref.~\cite{frsim}, the condition for high curvature in high-density regions in $f(R)$ gravity is closely related to the inequality
\begin{equation}
c^2\left|\delta f_R\right| \leq \left|-\frac{\phi}{2}\right|\leq |-\frac{2}{3}\phi_N|\label{strongthinshell}\quad.
\end{equation}
Here, in addition to our previous numerical study~\cite{frsim}, we also provide a strict mathematical proof of the above inequality, making use of the Green's function. The details can be found in Appendix~\ref{appA}. Based on Eq.~(\ref{strongthinshell}), a necessary condition for the high-curvature in high-density regions requires that the depth of the local potential well $|-\phi|$ is close to or above the value of the background field $|-\phi|\gtrsim2c^2|\bar{f}_R|$~\cite{HuS,frsim}; a sufficient condition for the low-curvature in high-density regions is that $|-\phi|\ll2c^2|\bar{f}_R|$. If expressed in terms of the Newtonian potential $|-\phi_N|$, $|-\phi_N|\gtrsim\frac{3}{2}c^2|\bar{f}_R|$ is a weaker necessary condition for the high-curvature but $|-\phi_N|\ll\frac{3}{2}c^2|\bar{f}_R|$ is a stronger sufficient condition for the low-curvature in high-density regions.

\subsection{Cosmological simulations of $f(R)$ gravity}

Our cosmological simulations are performed using the {\sc ecosmog} code ~\cite{ECOSMOG} which itself is based on the $N$-body code {\sc ramses}~\cite{RAMSES}. The box size is
$L_{\rm box}=150 h^{-1}{\rm Mpc}$. The number of particles is $N=256^3$. The cosmological parameters are $\Omega_b^0=0.049, \Omega_c^0=0.267, \Omega_d^0=0.684, h=0.671, n_s=0.962$, and $\sigma_8=0.834$, which are the Planck 2013~\cite{planck1} best-fit values for the standard $\Lambda$CDM model. We use the {\sc Mpgrafic} package~\cite{inicon} to generate initial conditions at $z=49$. We choose the parameter $f_{R0}$ for our $f(R)$ model as $f_{R0}=-10^{-6}$,$f_{R0}=-10^{-5}$,$f_{R0}=-10^{-4}$. In addition to the $f(R)$ simulations, we also implement a suite of $\Lambda$CDM simulations with the same box size, cosmological parameters and initial conditions as control.

\section{Power spectrum\label{power}}
\subsection{The impact of background field $f_{R}$}

As we have seen above,  the scalar field $f_{R}$ enters the field equations Eq.~(\ref{poissonfr}) and Eq.~(\ref{frpoisson}) via $R(f_R)$. Thus, to numerically solve the equations, we need to have analytical expressions of $R(f_R)$, which can in principle be obtained by inverting $f_R(R)$, which itself can be derived from Eq.~(\ref{fr}). However, Eq.~(\ref{fr}) is complicated and it is difficult to extract $R(f_R)$ analytically. Following \cite{frsim} we will instead use fitting formulae for $R(f_{R})$.

\begin{figure}
\includegraphics[width=3.5in,height=3.5in]{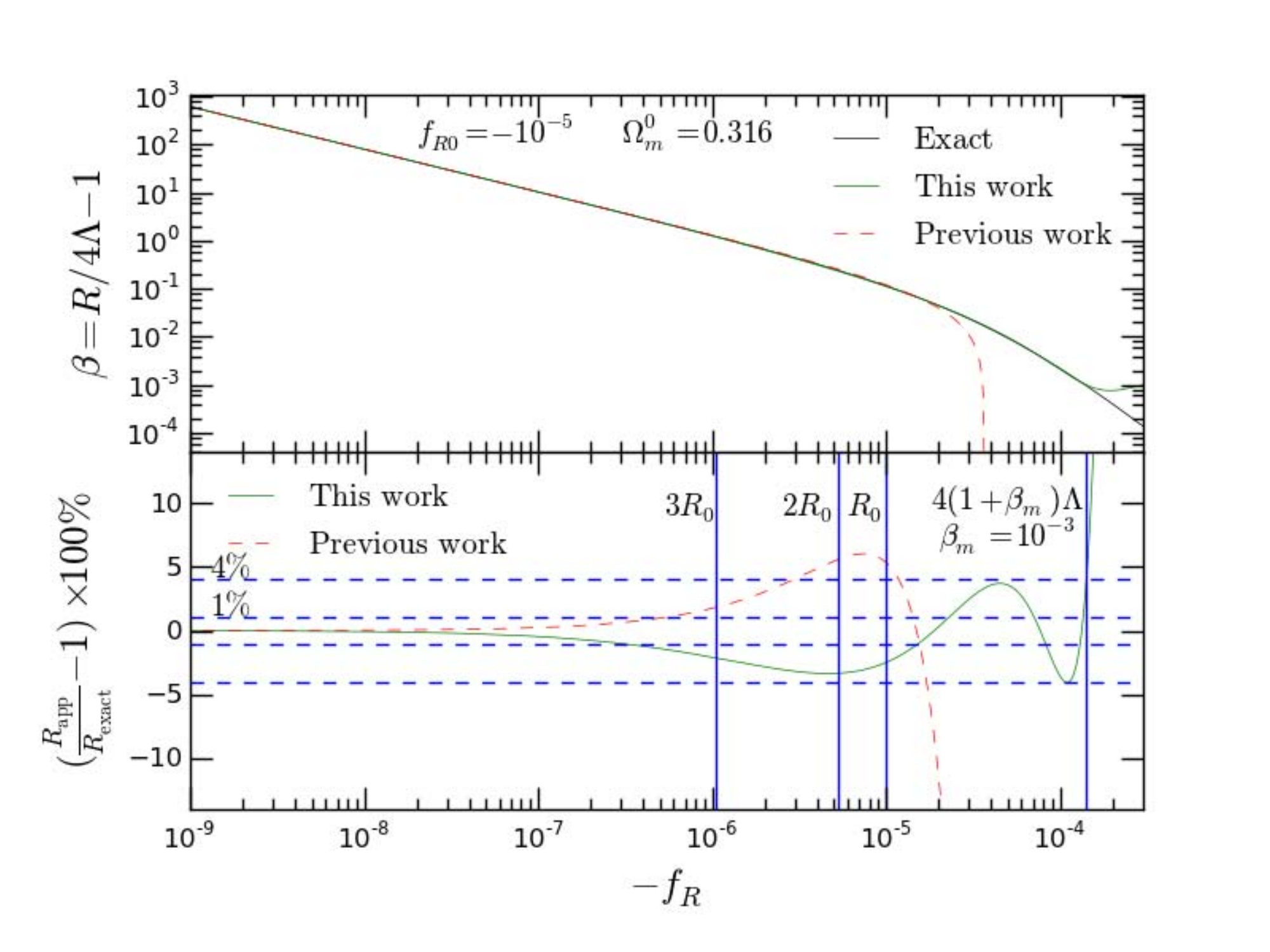}
\caption{Upper panel: $\beta = R(f_R)/4\Lambda -1$ as a function of $f_R$. It is clear that the accuracy of the fitting formula used in this work has been improved significantly particularly for small $R$. Lower panel: The error of fitting formulas. When $R\gtrsim 10R_0$, the error is below $1\%$.  When $R\sim R_0$, the error is around $2\%$. The error is always less than $4\%$ for $R>4(1+\beta_m)\Lambda$, where $\beta_m=10^{-3}$. }\label{fitting}
\end{figure}

%Since the background field $\bar{f}_R$ is very important for $f(R)$ simulations, i
In this work, we propose an improved fitting formula of $R(f_R)$ for our $f(R)$ model
\begin{equation}
R(f_R)=12\Omega_d^0H_0^2+3\Omega_m^0\left[\left(\frac{D}{f_R}\right)^{\frac{1}{p_{+}}}-\eta e^{-\left(\frac{f_R}{D}\right)^{\alpha}}\right]H_0^2\quad,
\end{equation}
where $\alpha$ and $\eta$ are fitting parameters. Fig.~\ref{fitting} shows the accuracy of this improved fitting formula compared to the exact expression obtained from Eq.~(\ref{fr}) and the one used in%in our previous work
~\cite{frsim}. $\eta$ and $\alpha$ depend on $\Omega_m$, and their values here are taken as $\eta=1.3038$, $\alpha = 0.3733$ for $\Omega_m=0.316$. However, they are independent of $D$ and $f_{R0}$. Fig.~\ref{fitting} shows that, when $R\gtrsim 10R_0$ the error of the fitting formula is well below $1\%$, when $R\sim R_0$ the error is around $2\%$, and the overall error is always smaller than $4\%$ for $R>4(1+\beta_m)\Lambda$ where $\beta_m=10^{-3}$.

\begin{figure}
\includegraphics[width=3.5in,height=3.5in]{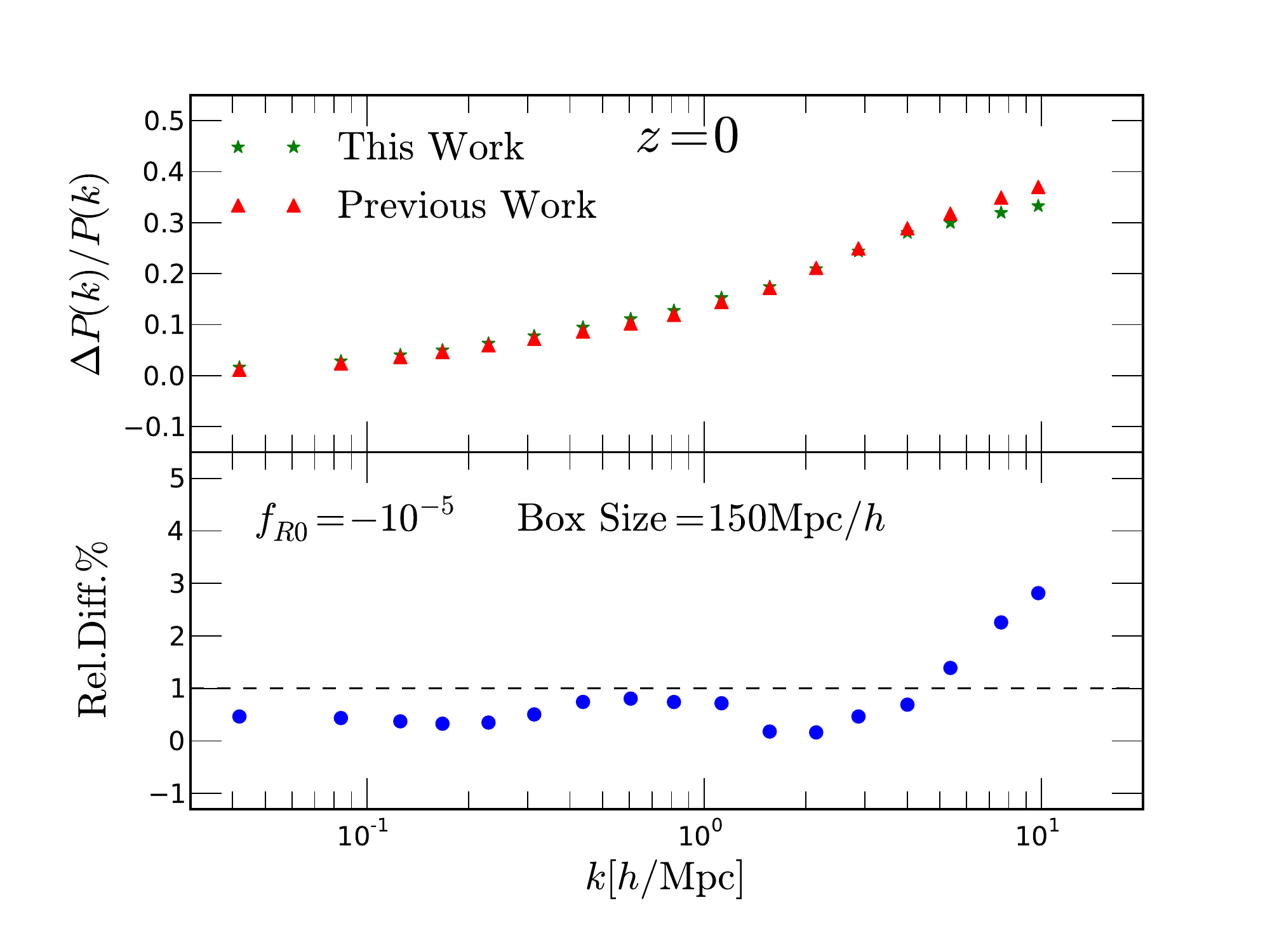}
\caption{The power spectra measured from our previous and current simulations using {\sc powmes}. The relative difference is well below $1\%$ for $k< 5 h/{\rm Mpc}$, and the difference on small scales $k>5 h/{\rm Mpc}$ is less than $3\%$.}\label{pk}
\end{figure}

To examine the impact of the accuracy of the fitting for $R(f_R)$  on matter power spectra, we implement a test simulation, using the same initial conditions as in our previous work~\cite{frsim}. The number of particles in the test is  $N=256^3$ and the box size is $L_{\rm box}=150 h^{-1}{\rm Mpc}$. We choose $f_{R0}=-10^{-5}$ for illustrative purposes since this model has the most complex shape of the power spectrum and most complicated screening situation at $z=0$. The numerical results are shown in Fig.~\ref{pk}, in which the upper panel shows the relative difference between the power spectra of our $f(R)$ and the $\Lambda$CDM model
\begin{equation}
\Delta P/P=P_{f(R)}/P_{\Lambda \rm{CDM}}-1\quad,\label{deltaP}
\end{equation}
at $z=0$, measured by using the {\sc powmes} \cite{POWMES} code. The lower panel shows the relative difference on the power spectra between this work and our previous work $$\left |\frac{P_{\rm{previous\,work}}}{P_{\rm{this\,work}}}-1\right |\times 100\%\quad.$$
We find a good agreement on the measured power spectra: the relative difference is below $1\%$ up to $k\sim 5 h/{\rm Mpc}$, and even on the smallest scales probed by this simulation ($k\sim 10 h/{\rm Mpc}$) it is less than $3\%$. We therefore conclude that the error induced by the fitting formula of $R(f_R)$ has been controlled within a fairly reasonable range.

\subsection{Power spectra of scalar fields}

To achieve a high spatial resolution in simulations, the {\sc ramses} code employs the adaptive mesh refinement (AMR) technique. %, which has already been implemented in the {\sc ramses} code.
The simulation box is initially covered by a domain mesh with a fixed resolution. Each cell of the domain mesh is hierarchically refined according to some pre-defined criteria (e.g.,  a density threshold) during the %process of
simulation. Although the domain mesh resolution in our simulation is only $256^3$, the highest resolution of the refined cells could be as high as $(2^{15})^3=32768^3$. The physical quantities (e.g., densities, potentials) are assigned on the AMR grid, at the centers of the cells, during the simulation.

In this work, we make use of this AMR grid to measure the power spectra of various scalar fields ($\delta\rho$, $\delta\rho_{\rm eff}$, $\phi$, $\phi_N$, $c^2\delta f_R$). Our method is similar to measuring the power spectrum of gas pressure in hydrodynamical simulations~\cite{RAMSES}. Unlike the density field, other scalar fields such as the gas pressure and various potentials can not be easily sampled by particles without bias. The AMR grid is therefore a natural choice for this work. In order to do this, we record the values of these fields at different levels of the AMR grid for each snapshot. However, we only use the leaf cells (which are not refined) at each snapshot. The {leaf} cells can seamlessly cover the whole simulation domain.

As discussed in Appendix~\ref{sect:alias}, the power spectrum of a continuous scalar field $u(\vec{x})$, where $u$ represents any one of $\delta\rho$, $\delta\rho_{\rm eff}$, $\phi$, $\phi_N$ and $c^2\delta f_R$, is defined by
\begin{equation}
\left\langle U(\vec{k}_1)U(\vec{k}_2)^*\right\rangle \equiv(2\pi)^3\delta^D(\vec{k}_1-\vec{k}_2)P_u(k)\, ,
\end{equation}
where $U(\vec{k})$ is the Fourier transform of $u(\vec{x})$ and $\delta^D$ is the Dirac delta function. Although this definition %of power spectrum
assumes a continuous scalar field, in practice the power spectrum can only be measured by discrete samplings.
Usually, the scalar field is sampled on a set of regularly spaced grid points and then analyzed by using the Fast Fourier Transform (FFT). As explained in Appendix~\ref{sect:alias}, the biggest problem of using FFT to measure the power spectrum is the aliasing effect, namely the discrete Fourier transform does not give the power spectrum of the scalar field but a sum of aliases of the continuous Fourier transform of the scalar field
\begin{equation}
P^{FFT}_u(\vec{k})=\sum_{\hat{n}}P_u(\vec{k}+2k_N\hat{n})\quad,
\end{equation}
where $k_N=\frac{\pi N_g}{L}$ is the Nyquist frequency, $L$ is the box size, $N_g$ is the number of cells in one dimension of the FFT mesh, and $\hat{n}$ is a position vector which indicates the alias.
Without prior knowledge of the power spectrum (e.g., its shape), it is impossible to remove the aliasing effect from the true signal. However, it is possible to minimize such aliasing effects by increasing the resolution of the %sampling of
FFT mesh. If the Nyquist frequency is high enough, the alias can be separated from the true signal in the main region $0<k<2k_N$ and so its effect can be minimized.

In this work, we assign %the value of the scalar field
the values of scalar fields to the FFT grid without smoothing
$$u_{\vec{x}_g}=u(\vec{x})|_{\vec{x}=\vec{x}_g}\, ,$$
where $\vec{x}_g$ represents the positions of the FFT grid points, and $u_{\vec{x}_g}$ is the sampled signal on those points. This is equivalent to interpolating the scalar field $u(\vec{x})$ from the leaf AMR cells to the FFT grid points $\vec{x}_g$ directly. In detail:
\begin{itemize}
\item If the {leaf} AMR cell is coarser than or the same coarseness as the FFT cell, we assign the value of the scalar field in the AMR cell to all of the FFT grid points that are contained within the AMR cell.
\item If the {leaf} AMR cell is finer than the FFT cell, we interpolate values from the eight nearest surrounding AMR grid points to the FFT grid.
\end{itemize}
This method enables us to measure the power spectra of any scalar field, and we call it the AMR-FFT method for convenience. Before presenting our results, we will perform several tests on this method.

\begin{figure}
\includegraphics[width=3.5in,height=3.5in]{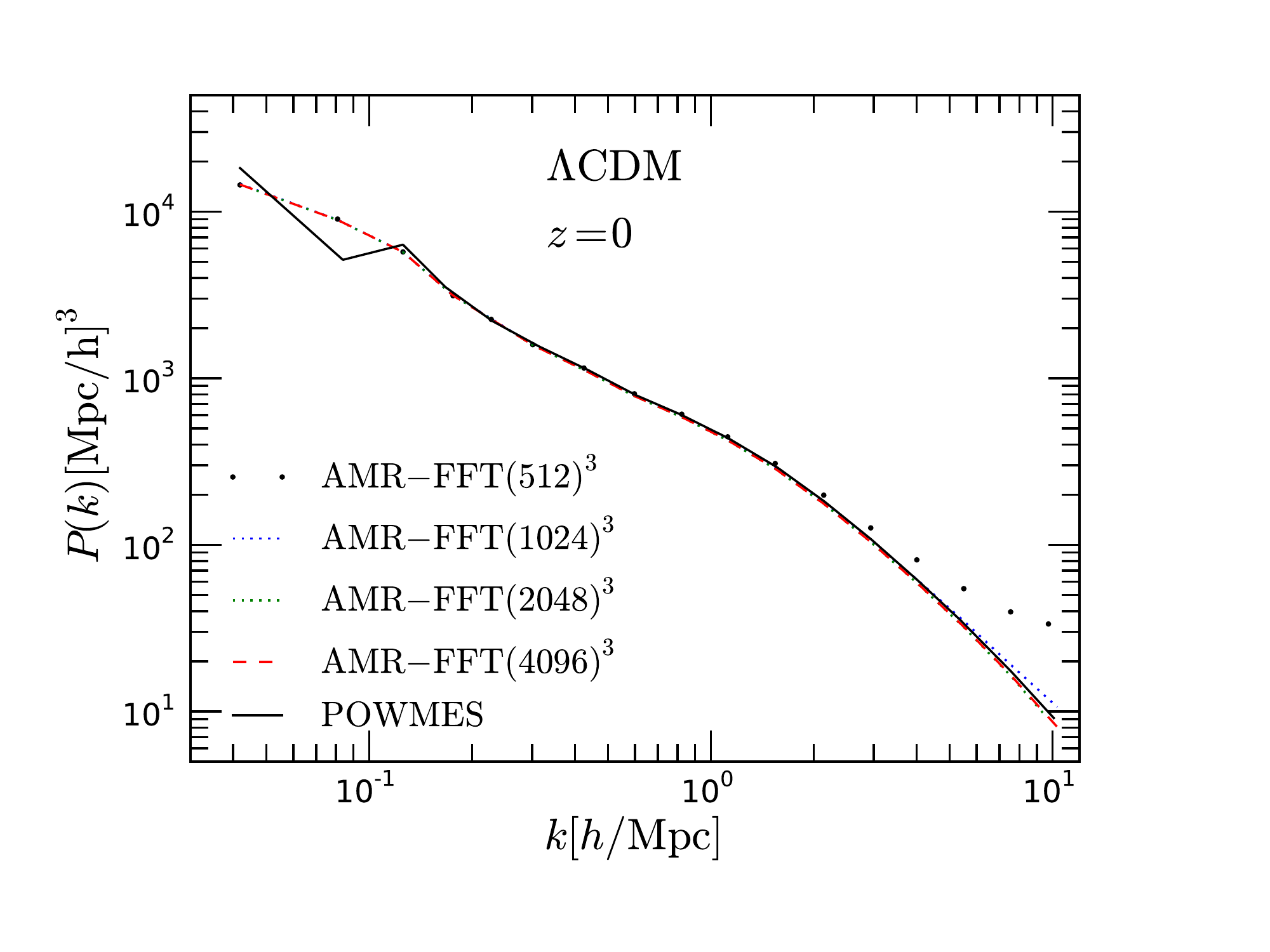}
\caption{The power spectrum of the $\Lambda$CDM model measured from the {\sc powmes} code (the solid line) and our AMR-FFT method with different resolutions.
Our highest resolution measurement $4096^3$ agrees with the {\sc powmes} code very well out to $k\sim 10 h/{\rm Mpc}$. }\label{LCDMPK}
\end{figure}

\begin{figure}
\includegraphics[width=3.5in,height=3.5in]{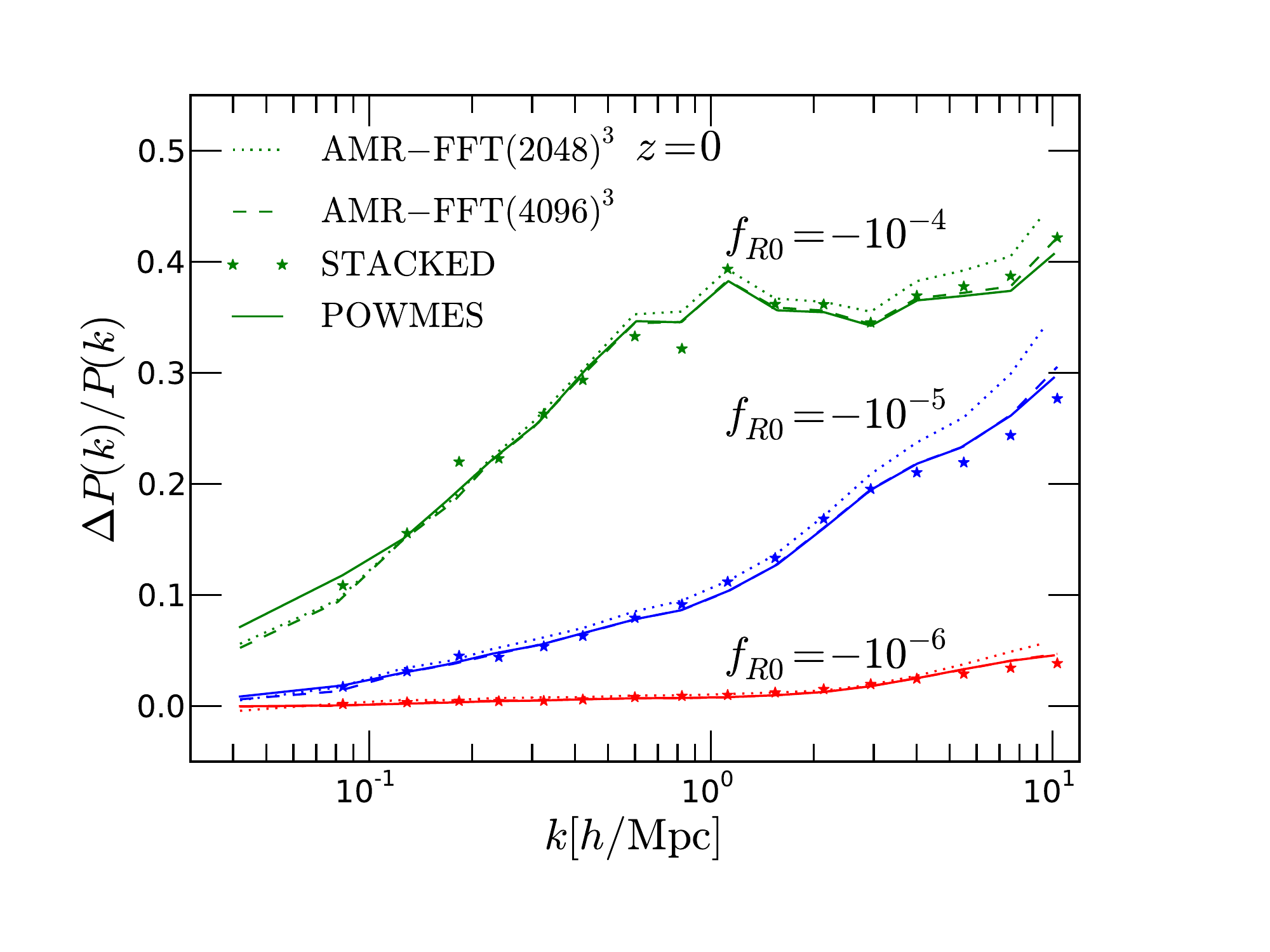}
\caption{The relative differences of the conventional dark matter power spectra with respect to the $\Lambda$CDM model for different $f(R)$ models. The $2048^3$ FFT measurements yield about $5\%$ relative differences on small scales (dotted lines), while the high resolution measurements $4096^3$ (dashed lines) agree with {\sc powmes}  (solid lines) better than $1\%$. The stars indicate the results obtained by stacking density fields. }\label{frPK}
\end{figure}

Our first test is about the accuracy of the AMR-FFT method when applied to the power spectrum of the dark matter density field, $P_m(k)$.
$P_m(k)$ can be measured to a high accuracy directly from the dark matter particles. We adopt the results from the {\sc powmes} code as a control. {\sc powmes} is based on a method of Taylor expansion of trigonometric functions, and can yield an unbiased and nearly alias-free estimation of %the dark matter power spectrum.
$P_m(k)$. For the purpose of comparison, we only focus on one realization, and choose $f(R)$ models with different parameters as well as the $\Lambda$CDM model. These simulations share the same initial conditions. In Fig.~\ref{LCDMPK}, we show the dark matter power spectra for the $\Lambda$CDM model measured using our AMR-FFT method with different resolutions of the FFT grid. As shown in Fig.~\ref{LCDMPK}, the accuracy of the AMR-FFT method depends strongly upon the resolution of the FFT grid. A low resolution measurement such as $512^3$ ($k_N/4 \sim 2.68 h/{\rm Mpc}$) gives very poor accuracy on the power spectrum on small scales $k>k_N/4$. However, with a much higher resolution, such as $4096^3$ ($k_N/4 \sim 20 h/{\rm Mpc}$),
this method agrees with the {\sc powmes} code very well. %Next, in order to quantitatively show the accuracy of our measurements,
Further, we also show in Fig.~\ref{frPK} the quantity $\Delta P/P$ as defined in Eq.~(\ref{deltaP}) for different $f(R)$ models. Again, using the $4096^3$ FFT grid, the AMR-FFT method agrees with {\sc powmes} better than $1\%$ out to $k\sim 10 h/{\rm Mpc}$.

In Ref.~\cite{PKmeasure}, a multi-grid method which is based on a hierarchy of nested cubic Cartesian grids is proposed in order to save the use of memory in the computation of the FFT. The power spectra can be measured by dividing the volume of a box into small cubes and then stacking the small cubes into a co-added density field. Instead of using a single high resolution FFT, only a few times of relatively low resolution FFT are needed in this method. The final power spectrum is obtained by combining the different ``band power" measurements obtained from different volumes of the stacked density fields~\cite{PKmeasure}. In Fig.~\ref{frPK}, we also present the power spectrum measured using this method. However, as shown in Fig.~\ref{frPK}, the stacked density yields about $3\%$ error on the power spectrum relative to {\sc powmes}. In order to get more accurate results, we therefore do not use this method in this work.

\begin{figure}
\includegraphics[width=3.5in,height=3.5in]{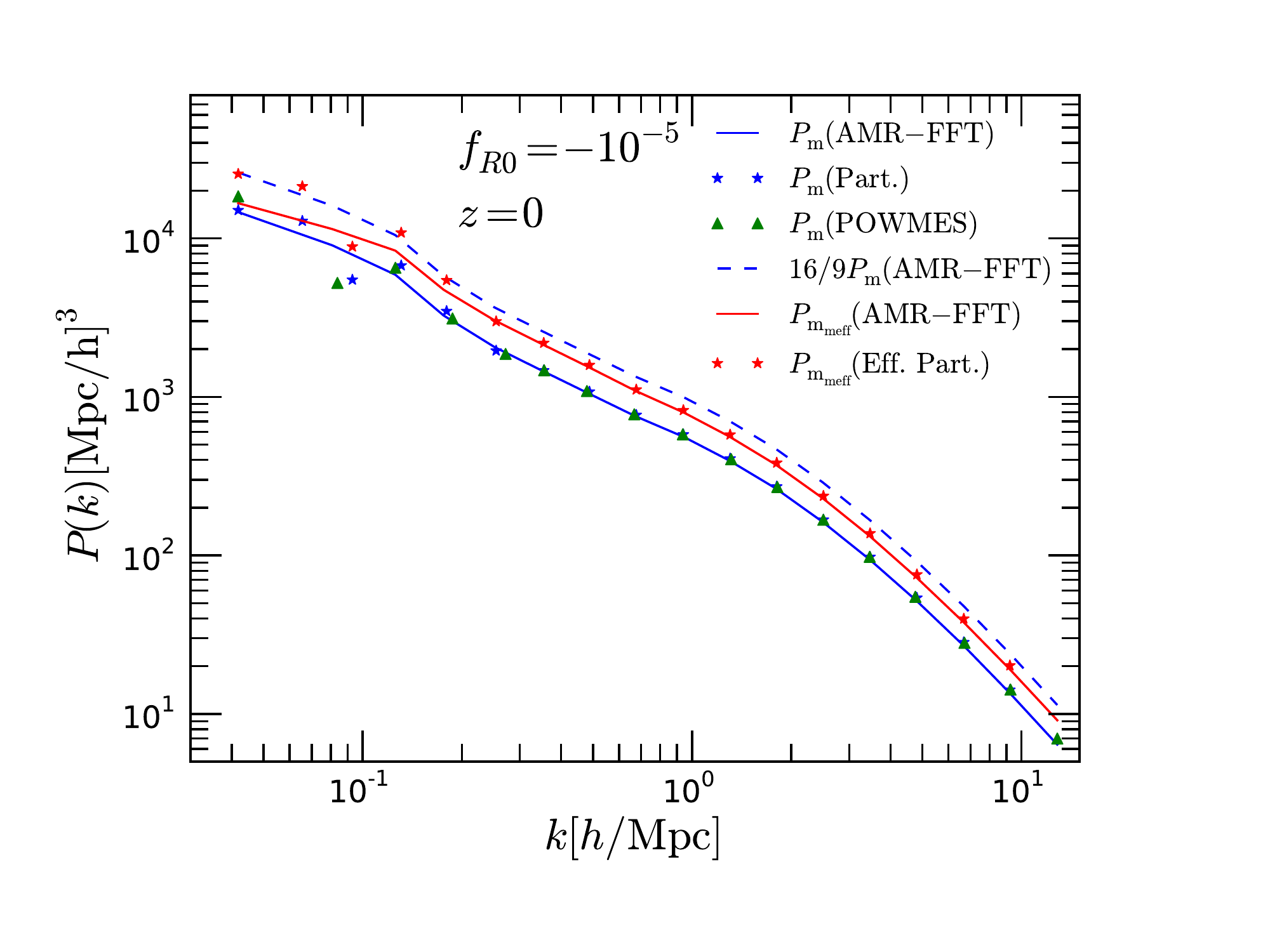}
\caption{The power spectrum of the effective density field measured by our AMR-FFT method (red solid line). The red stars indicate the power spectrum of the mass weighted particles. For comparison, the power spectra of the dark matter density field are also shown. Our code (blue stars), the POWMES code (green triangles) and the AMR-FFT method (blue solid line) agree well on the power spectrum of the dark matter density field $P_m$. }\label{EffPK}
\end{figure}

Our second test concerns the measurement of the power spectrum of the effective density field, $\delta_{\rm eff}$, which is defined by
\begin{equation}
\delta_{\rm eff}\equiv\frac{\delta \rho_{m_{\rm eff}}}{\bar{\rho}_m}=\frac{G_{\rm eff}}{G}\frac{\delta \rho_m}{\bar{\rho}_m}\quad.
\end{equation}
Using the AMR-FFT method, the power spectrum of $\delta_{\rm eff}$ can be measured in a similar way to that of the density field. For illustrative purposes, we only focus on the $f(R)$ model with $f_{R0}=-10^{-5}$, and the results are shown in Fig.~\ref{EffPK}. As a comparison, we also present the dark matter power spectrum $P_m(k)$. Compared with %the dark matter power spectrum
$P_m(k)$, the effective power spectrum $P_{m_{\rm eff}}(k)$ is enhanced due to the enhancement of gravity. However, $P_{m_{\rm eff}}(k)$ should not exceed $\frac{16}{9}P_m(k)$ since the maximal enhancement of gravity is $1/3$. This upper limit is indicated by the dashed line in Fig.~\ref{EffPK}.

To further test the validity and accuracy of our AMR-FFT method on the power spectrum of the effective density field, instead of using the regular grid for sampling the effective density field, we sample the effective density field at the positions of dark matter particles. We treat the effective density field as discrete mass-weighted dark matter particles
\begin{equation}
\begin{split}
\rho_{m_{\rm eff}}(\vec{x})&\approx\frac{G_{\rm eff}}{G} \rho_m(\vec{x}) = \frac{G_{\rm eff}}{G}\sum_{i}m \delta^D(\vec{x}-\vec{x}_i)\\
&=\sum_{i} \frac{G_{\rm eff}(\vec{x}_i)m}{G} \delta^D(\vec{x}-\vec{x}_i)\\
&=\sum_{i} m_{\rm eff}(\vec{x}_i) \delta^D(\vec{x}-\vec{x}_i)\quad,\label{meff}
\end{split}
\end{equation}
in which $m$ is the true mass of particles and $m_{\rm eff}\equiv\frac{G_{\rm eff}}{G}m$ is the effective mass. Although the discrete particles induce shot noise, the Fourier transform of the density field, in principle, can be evaluated accurately by a direct sum over the modes of the Fourier transform of each particle
\begin{equation}
\rho_{m_{\rm eff}}(\vec{k})=\sum_{i}m_{\rm eff}(\vec{x}_i)e^{-i\vec{k}\cdot\vec{x}_i}\quad.\label{fourier}\\
\end{equation}
However, this method is numerically unfeasible for N-body simulations since they usually contain a large number of particles and  Eq.~(\ref{fourier}) does not have a fast algorithm like the FFT due to the irregular distribution of points. In practice, in order to improve the efficiency, the power spectrum of discrete particles can be measured by assigning particles to a regular grid and then analyzing them using the FFT. The assignment of particles is equivalent to smoothing the underlying density field and then putting the averaged values on an FFT grid. This smoothing effect, however, can be exactly corrected afterwards. Furthermore, with the aid of further correction strategies (e.g. Ref.~\cite{Jing}), the aliasing effect can be significantly suppressed and a reasonable accuracy of the power spectrum can be obtained with less computational expense. Although this method is not fully free of bias and aliasing in general, it provides an alternative way to our AMR-FFT method for measuring the power spectrum of the effective density field. We therefore use it as a cross check of our measurement %on the power spectrum of the effective density field
of $P_{m_{\rm eff}}$.

In high density regions $\delta\gg 1$, the mass weighted sampling should be unbiased and the power spectrum of the mass weighted particles should be close to the power spectrum of the effective density field. Since the {\sc powmes} code has not been tested for mass weighted particles, we instead use our own code. Our code follows the method as proposed in Ref.~\cite{Jing}. We remove the aliasing effect by assuming a power law $P_{m_{\rm eff}} \propto k^{\alpha}$ for the power spectrum on large $k$, where $\alpha$ is a fitting parameter. Following Ref.~\cite{Jing}, we also correct the shot noise induced by the discrete sampling of particles and the smoothing window function of the particle assignment. The results are shown as red stars in Fig.~\ref{EffPK}. As a demonstration of our code, we also present the measurement of the dark matter power spectra $P_m$ from the dark matter particles using our code. As is indicated by the blue stars, our code yields the same results as the {\sc powmes} code and the AMR-FFT method. As for the power spectrum of the effective density field, on small scales the mass weighted sampling agrees very well with the results from our AMR-FFT method. However, on large scales, there are some deviations. This can be expected because the sampling of mass weighted dark matter particles is biased in low density regions for the effective density field. In these regions, Eq.~(\ref{meff}) is less accurate and, moreover, there might be few or no dark matter particles in low density regions while the effective density field is continuous and can not be zero.

\begin{figure}
\includegraphics[width=3.5in,height=3.5in]{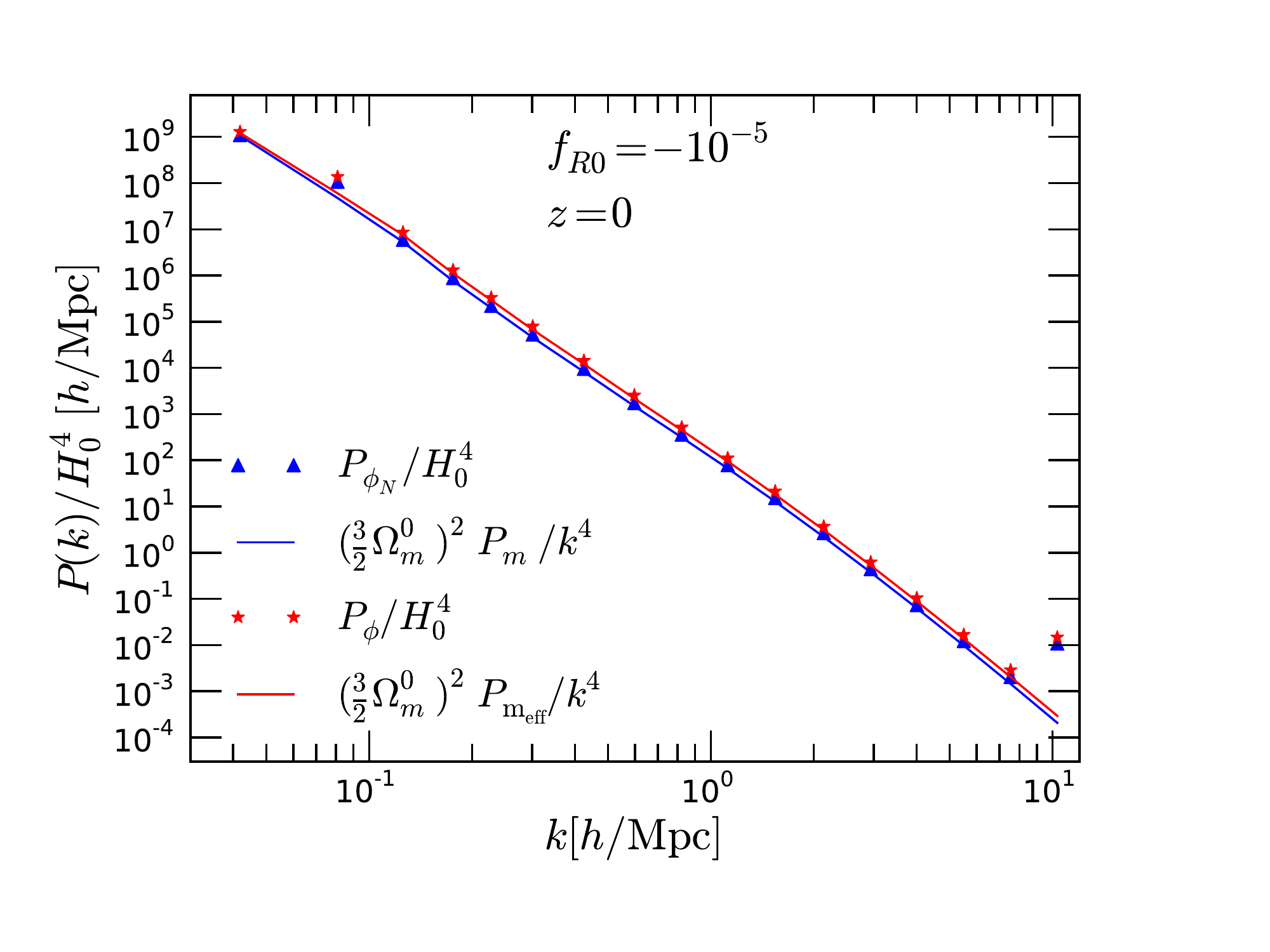}
\caption{The power spectra of $\phi_N$ (blue triangles) and $\phi$ (red stars) measured using our AMR-FFT method as compared with the spectra derived from the density fields (solid lines).}\label{potential}
\end{figure}

In addition to the above tests, we also make several consistency tests. From the Poisson equations, Eq.~(\ref{poissonN}) and Eq.~(\ref{eff_poisson}), the power spectra of potentials and density fields should follow the relations
\begin{equation}
\begin{split}
P_{\phi}&=\left(\frac{3}{2}\Omega_m^0\right)^2\frac{H_0^4}{a^2k^4}P_{m_{\rm eff}}\quad,\\
P_{\phi_N}&=\left(\frac{3}{2}\Omega_m^0\right)^2\frac{H_0^4}{a^2k^4}P_m\quad.
\end{split}
\end{equation}
In our simulations, the potentials are obtained by using a relaxation method in real space~\cite{ECOSMOG}, which is different from the FFT method. It is therefore important to check the consistency of the above relations. In Fig.~\ref{potential}, we show the directly measured power spectra of $\phi$, $\phi_N$ and the corresponding power spectra derived from the density fields (the right-hand sides of the above equations). Most strikingly, the numerical results agree very well over $12$ orders of magnitude. This serves not only as a test of the consistency, but also as a check of the AMR-FFT method employed in this work.


\begin{figure}
\includegraphics[width=3.5in,height=3.5in]{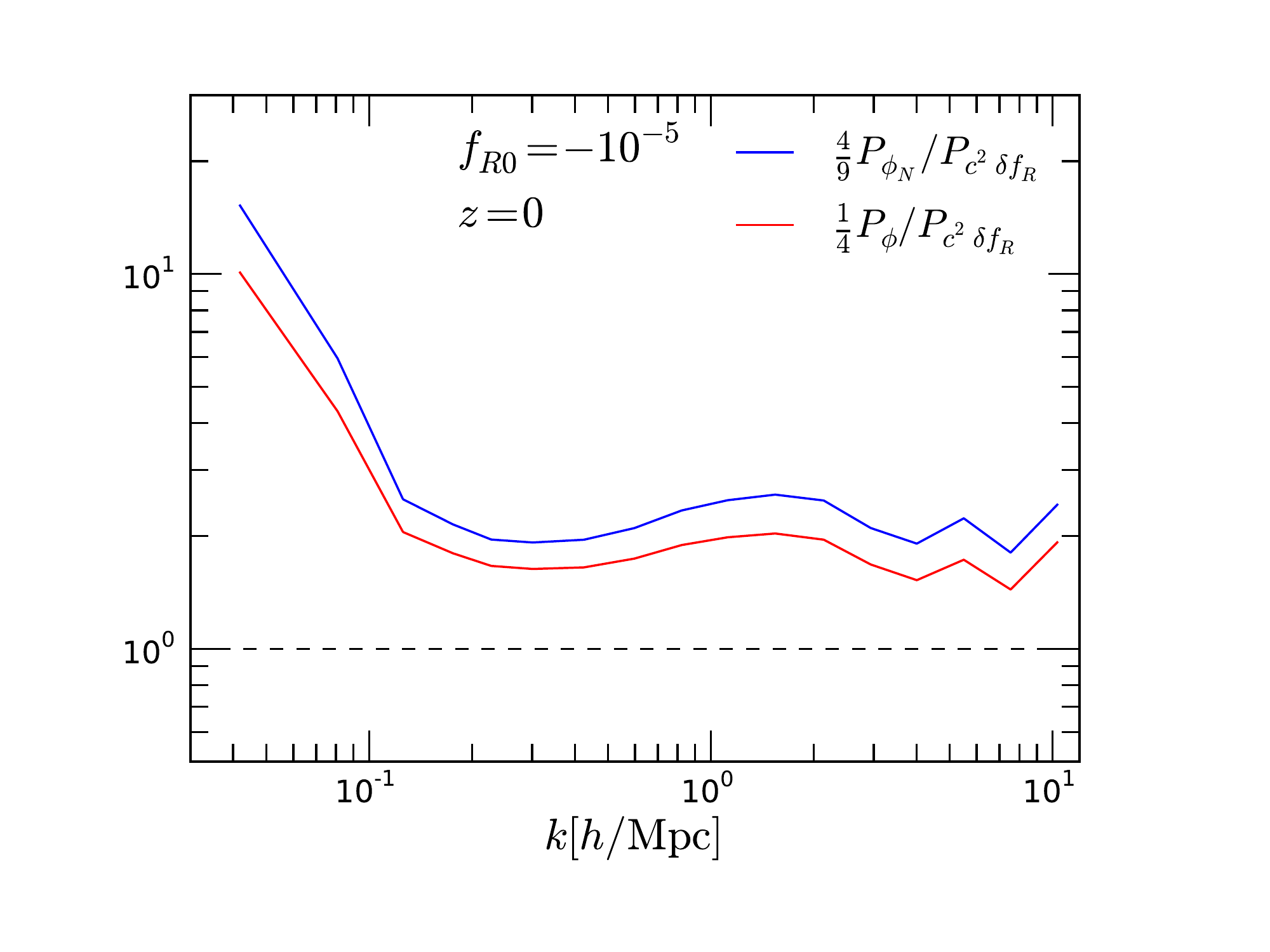}
\caption{The ratio of $\frac{1}{4}P_{\phi}/ P_{c^2\delta f_R}$ and $\frac{4}{9} P_{\phi_N}/ P_{c^2\delta f_R}$ as a function of $k$. }\label{inequalities}
\end{figure}

Finally, besides the above consistency relations, according to Eq.~(\ref{strongthinshell}), the power spectra of the scalar fields should also satisfy the following inequalities
\begin{equation}
P_{c^2\delta f_R}(k)\leq \frac{1}{4} P_{\phi}(k) \leq \frac{4}{9} P_{\phi_N}(k) \quad.\label{ineq}
\end{equation}
In Fig.~\ref{inequalities}, we show the ratio of $\frac{1}{4}P_{\phi}/ P_{c^2\delta f_R}$ and $\frac{4}{9} P_{\phi_N}/ P_{c^2\delta f_R}$ as a function of $k$, from which we can see that indeed Eq.~(\ref{ineq}) holds on scales from $k\sim 0.06 h/{\rm Mpc}$ to $k\sim 10 h/{\rm Mpc}$.


\subsection{Power spectra of the effective density field}

The above tests give us confidence in the validity and accuracy of our AMR-FFT method. In this subsection, we present the power spectra of the effective density field measured by averaging over five realizations of simulations. %All the power spectra, in this subsection, are measured by using our
Following the results of those tests, we use a $4096^3$ FFT grid in this subsection to measure the power spectra of density fields and various potentials.

\begin{figure}
\includegraphics[width=3.5in,height=3.5in]{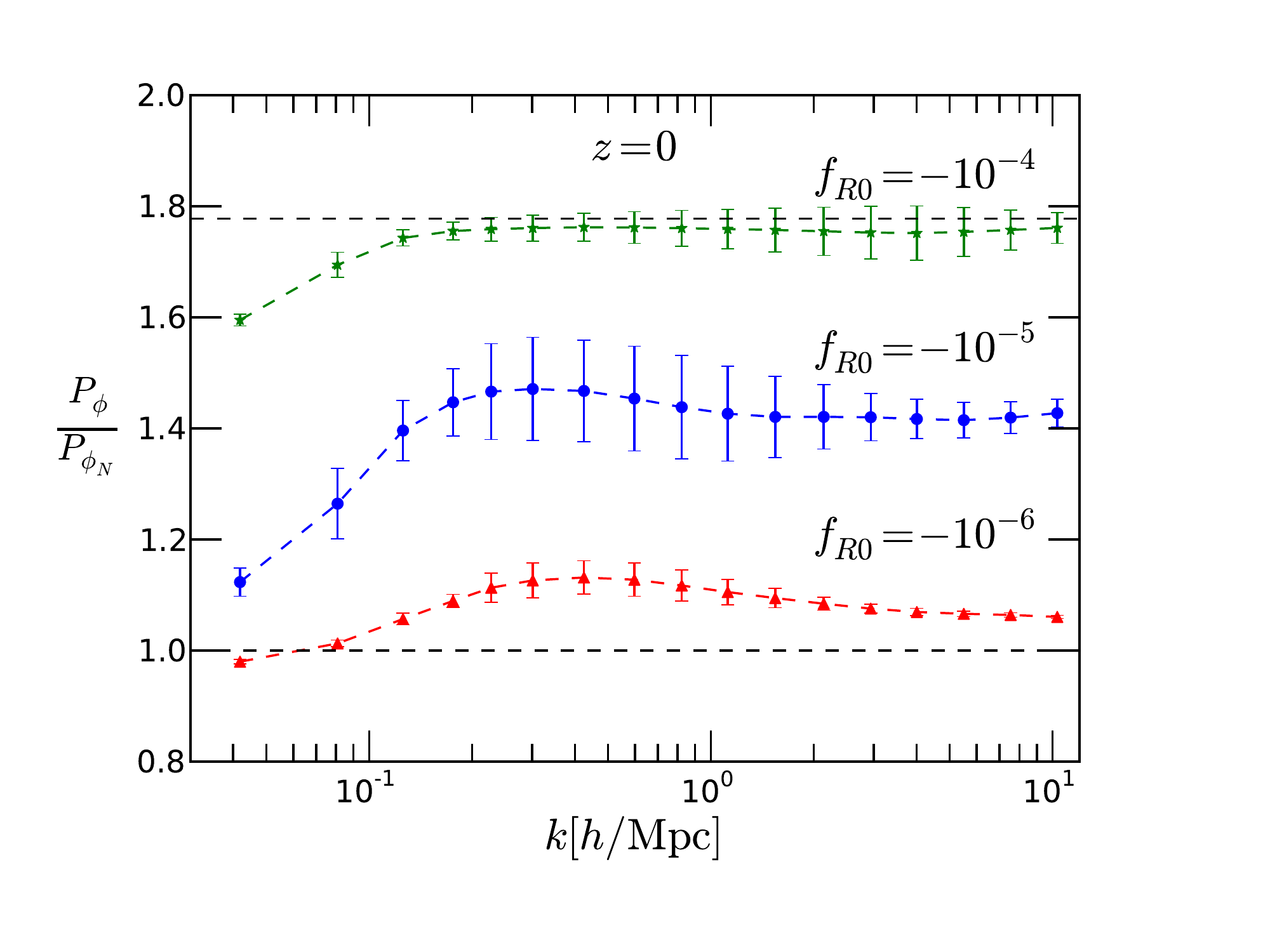}
\caption{The ratio of the power spectra $P_{\phi}/P_{\phi_N}=P_{m_{\rm eff}}/P_m$ as a function of $k$. The lower and upper dashed lines indicate the value of $1$ and $16/9$, respectively. The ratio of the spectra should satisfy $1<P_{\phi}/P_{\phi_N}<16/9$. For $f_{R0}=-10^{-4}$, due to the lack of screening, $P_{\phi}/P_{\phi_N}$ is very close to $16/9$. For $f_{R0}=-10^{-6}$, the screening mechanism works efficiently and so $P_{\phi}/P_{\phi_N}$ is close to $1$. }\label{PPNratio}
\end{figure}

In Fig.~\ref{PPNratio}, we show the ratio of the power spectra of the two potentials
\begin{equation}
\frac{P_{\phi}}{P_{\phi_N}}=\frac{P_{m_{\rm eff}}}{P_m}\quad.
\end{equation}
As indicated in Fig.~\ref{PPNratio}, the ratio satisfies $1<P_{\phi}/P_{\phi_N}<16/9$ for all simulated $f(R)$ models, consistent with the prediction by Eq.~(\ref{ineq}). For $f_{R0}=-10^{-4}$, due to the lack of an efficient Chameleon screening, the gravitational force is enhanced by $1/3$ and therefore $P_{\phi}/P_{\phi_N}$ is very close to $16/9$. On the other hand, for $f_{R0}=-10^{-6}$, the screening mechanism works efficiently and hence $P_{\phi}/P_{\phi_N}$ is very close to $1$. The situation for $f_{R0}=-10^{-5}$ is somewhere in between, as expected.%just in the medium between $f_{R0}=-10^{-4}$ and $f_{R0}=-10^{-6}$.

\begin{figure}
\includegraphics[width=3.5in,height=3.5in]{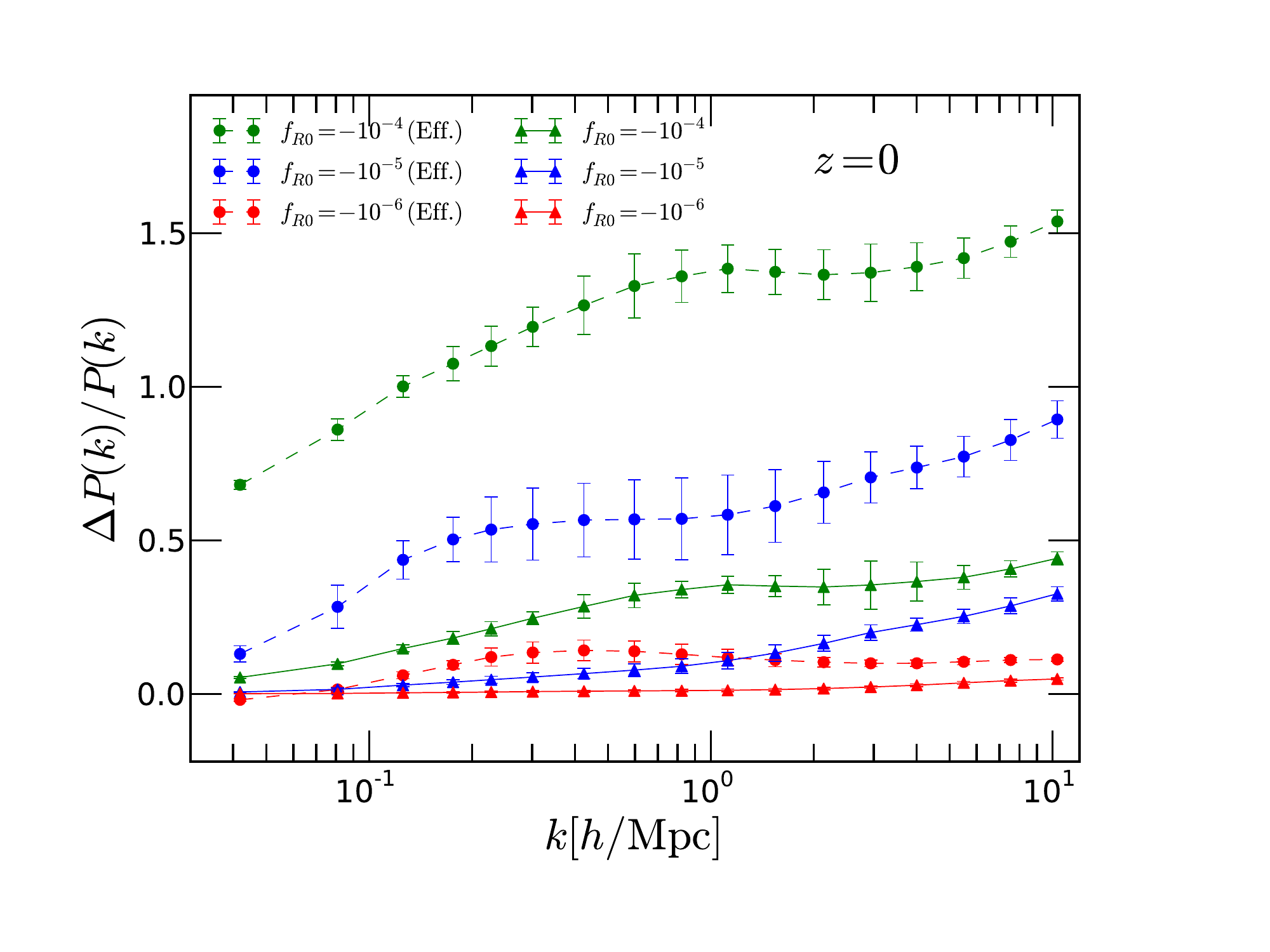}
\caption{The relative differences of power spectra with respect to the $\Lambda$CDM model for the dark matter density field (triangles with solid lines) and the effective density field (solid circles with dashed lines), respectively. The power spectra of the effective density fields $P_{m_{\rm eff}}$ show more significant deviations from the $\Lambda$CDM model than those of the dark matter density field $P_m$.}\label{deltaEFFPK}
\end{figure}

Next, we show the relative difference of the effective power spectra with respect to the $\Lambda$CDM model. In Fig.~\ref{deltaEFFPK}, we present $\Delta P_m/P_{\Lambda \rm{CDM}}$ and $\Delta P_{m_{\rm eff}}/P_{\Lambda \rm{CDM}}$ for the simulated $f(R)$ models. %which is defined by Eq.~(\ref{deltaP}).
%Most strikingly,
As expected, compared with the dark matter power spectra, the effective power spectra show more significant deviations from the $\Lambda$CDM model. For $f_{R0}=-10^{-4}$, $\Delta P_{m_{\rm eff}}/P_{\Lambda \rm{CDM}}$ %of the effective spectra
is about three times as large as $\Delta P_m/P_{\Lambda \rm{CDM}}$, which is because the former contains contributions from both the latter and the fact that $G_{\rm eff}$ is substantially enhanced compared with $G$. %that of the dark matter power spectra.
Even for $f_{R0}=-10^{-6}$, the effective power spectra show sizeable deviations (about $15\%$) from the $\Lambda$CDM model.
\subsection{The halo-halo power spectrum}
As shown in the previous subsection, the power spectra of $P_{\phi_N}$ are quite different from $P_{\phi}$ in $f(R)$ gravity. This is an interesting feature, which can be used to discriminate the $f(R)$ theory from other dark energy theories. To achieve this, however, we need to measure the power spectra of $P_{\phi_N}$ and $P_{\phi}$ independently. The lensing potential, $\phi_N$, can be measured directly from weak lensing surveys. Upcoming galaxy surveys such as the Euclid mission~\cite{Euclid} have the power to measure $P_{\phi_N}$ accurately on large scales.

The focus here is on how to measure $P_{\phi}$, which is, indeed, nontrivial. One possible method is to measure the galaxy cluster-cluster power spectrum. In fact, from Fig.~\ref{PPNratio}, it can be seen that the differences between $P_{\phi_N}$ and $P_{\phi}$ also exist on relatively large scales. $P_{\phi}$ on large scales can be probed by the cluster-cluster power spectrum. In practice, it is more convenient to work with the ratio $P_{hh}^{\rm Eff}/P_{hh}$ directly, where $P_{hh}$ is halo-halo power spectrum of standard halos. The advantage of using $P_{hh}^{\rm Eff}/P_{hh}$ is that the ratio is independent of the halo bias and is practically measurable. In observations, galaxy clusters can be observed using different methods such as x-ray observations, lensing and the thermal Sunyaev-Zeldovich effect~\cite{SZ}. A wealth of information about the clusters can be obtained from these surveys. For example, the true mass can be inferred from the lensing data or the gas mass-cluster mass scaling relation~\cite{gasfrac}; the effective mass can be estimated either by the temperature-effective mass scaling relation~\cite{gasscaling} or by the profiles of gas density and temperature in x-ray surveys. In practice, we can first identify the clusters in x-ray surveys and measure the effective mass of each cluster. Then we divide the clusters into different mass bins. We measure $P_{hh}^{\rm Eff}$ for each mass bin. Similarly, we can measure $P_{hh}$ for each mass bin as well. In the $\Lambda$CDM model, for a given mass bin the measurements of $P_{hh}^{\rm Eff}$ and $P_{hh}$ should be the same $P_{hh}^{\rm Eff}/P_{hh}=1$. However, in $f(R)$ gravity, $P_{hh}^{\rm Eff}$ and $P_{hh}$ can be different. There are two main reasons for the differences. The major one is due to mass calibration. Given a mass bin, the number of clusters with messes determined by effective mass will be quite different from the number determined by the true mass (see mass functions in Ref.\cite{effhalos}). The second reason is due to the positions of cluster centers. The centers of effective masses do not necessarily overlap with the true masses, especially for small clusters. Since the power spectrum encompasses the above two effects, it should be more useful than the mass function to test $f(R)$ gravity.

\begin{figure*}
\includegraphics[width=7in,height=3in]{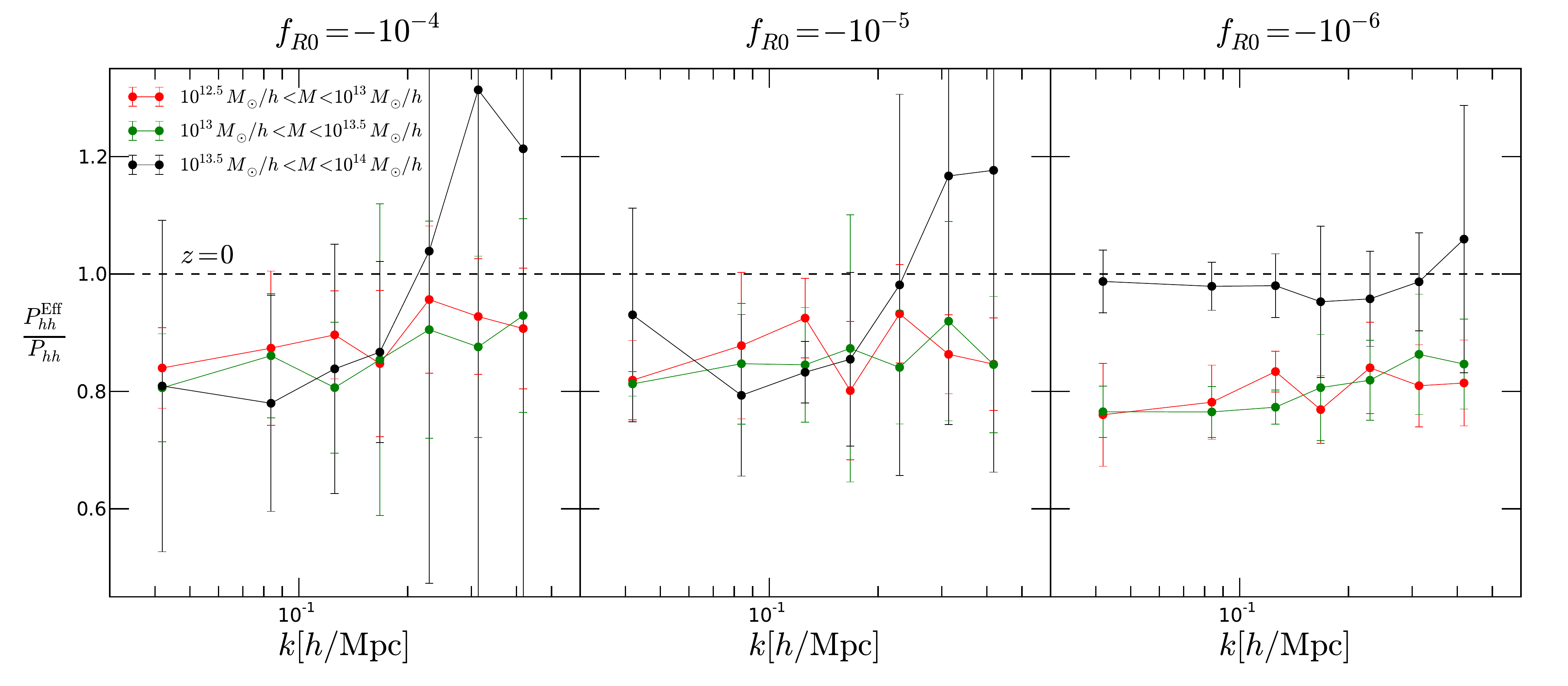}
\caption{The ratio of the effective halo power spectrum to the standard halo power spectrum for different $f(R)$ models. $P_{hh}^{\rm Eff}$ is measured from the effective catalog and $P_{hh}$ is measured from the standard catalog. Different colours represent different mass bins. In effective catalog, effective mass of halo is used and in standard halo catalog, the conventional mass is used. On relatively large scales ($k<0.2 h/{\rm Mpc}$), for $f_{R0}=-10^{-4}$ and $f_{R0}=-10^{-5}$, the ratio deviates from the $\Lambda$CDM prediction (the dashed line) at a level of $1\sigma$. For $f_{R0}=-10^{-6}$, due to the screening mechanism, the ratio for massive clusters ($M>10^{13.5}M_{\odot}/h$) is consistent with the $\Lambda$CDM prediction. However, for less massive clusters ($M<10^{13.5}M_{\odot}/h$), the deviations are over $3\sigma$.
}\label{haloPK}
\end{figure*}

Next, we test the above idea using simulations. We construct halo catalogs using a modified version of the {\sc Amiga} Halo Finder ({\sc Ahf})~\cite{AHF}. In halo-halo power spectrum $P_{hh}$, halo number density fields are represented by un-weighted particles. We therefore can use the {\sc powmes} code to measure the power spectrum. In Fig.~\ref{haloPK}, we present the ratio of the effective halo power spectrum to the standard halo power spectrum for different $f(R)$ models. Different colours represent different mass bins. In Fig.~\ref{haloPK}, note that shot noise has already been subtracted. On relatively large scales, the values of effective halo power spectra are less than those of the standard halos $P_{hh}^{\rm Eff}/P_{hh}<1$ for given mass bin. As shown in Fig.~\ref{haloPK}, on relatively large scales ($k<0.2 h/{\rm Mpc}$), for $f_{R0}=-10^{-4}$ and $f_{R0}=-10^{-5}$, the ratio $P_{hh}^{\rm Eff}/P_{hh}$ deviates from the $\Lambda$CDM prediction (the dashed line) at a level of almost $1\sigma$. For $f_{R0}=-10^{-6}$, due to the screening mechanism, the ratio for massive clusters ($M>10^{13.5}M_{\odot}/h$) is consistent with the $\Lambda$CDM prediction. However, for less massive clusters ($M<10^{13.5}M_{\odot}/h$), the deviations are over $3\sigma$. Further, it can also be found that for $f_{R0}=-10^{-4}$ and $f_{R0}=-10^{-5}$, the errors on the ratio are greater than for $f_{R0}=-10^{-6}$. This is because the effective power spectra have larger scatters for the former two cases than for $f_{R0}=-10^{-6}$ as shown in Fig.~\ref{deltaEFFPK}. The large errors on small scales ($k>0.2 h/{\rm Mpc}$) are due to shot noise. In particular, for the most massive clusters ($M>10^{13.5}M_{\odot}/h$), we only have a few hundred samples. The level of shot noise is quite high there. Finally, due to the limited box size and number of realizations, the simulations performed in this work tend to over estimate the errors. With a larger simulation box and more realizations, the error bars can be reduced and the deviations of $f(R)$ models from the $\Lambda$CDM model should be more clear. The results presented above therefore are very conservative.

\section{Summary and Discussion\label{conclusions}}
In this work, we have studied the non-linear power spectra of scalar fields in $f(R)$ gravity, using a suite of N-body simulations. We have illustrated in detail our AMR-FFT method for measuring the power spectra of scalar fields. Using this method, we have measured the power spectra of the potentials and the effective density fields. Our main results are summarized as follows:
\begin{itemize}
\item We have verified the inequality
 \begin{equation}
c^2\left|\delta f_R\right| \leq \left|-\frac{\phi}{2}\right|\leq |-\frac{2}{3}\phi_N|\quad,
\end{equation}
in Fourier space by comparing the power spectra of the potentials. As is discussed in Ref.~\cite{frsim}, the above inequality is closely related to the screening mechanism in $f(R)$ gravity. Its validity is important for predicting the screening theoretically.

\item We find that compared with the dark matter power spectra, the effective power spectra differ more significantly from the $\Lambda$CDM model. Even for $f_{R0}=-10^{-6}$, the effective power spectrum shows a sizeable signal of deviation.

\end{itemize}
We have tested that these conclusions are applicable to other $f(R)$ models as well. For Hu-Sawicki model~\cite{HuS}, we find similar results as presented in Ref.~\cite{Puchwein}.

Since the formation and clustering of galaxies in $f(R)$ gravity is more closely related to the effective density field rather than the dark matter density field itself, our findings indicate that the traditional statistics of the dark matter density field such as the power spectrum or equivalently the two-point correlation function may seriously under-predict the impact of modifications of gravity on the clustering of galaxies. However, it should be noted that although the power spectrum of the effective density field in $f(R)$ gravity is significantly different from that in the $\Lambda$CDM model, we should caution that these large deviations are likely to be highly degenerate with galaxy bias. Robust constraints on $f(R)$ gravity can thus only be obtained once we have a solid knowledge of the expected biasing of galaxies in $f(R)$ cosmologies.

Nevertheless, it is interesting to find that the power spectrum $P_{\phi_N}$ is quite different from $P_{\phi}$ in $f(R)$ gravity. This can be used to discriminate the theory from the $\Lambda$CDM model. We tested this idea by investigating the halo-halo power spectrum on relatively large scales. Galaxy clusters can be observed using different methods and a wealth of information about the clusters can be obtained. We found that probing the ratio $P_{hh}^{\rm Eff}/P_{hh}$ is an useful way to test $f(R)$ gravity. On relatively large scales ($k<0.2 h/{\rm Mpc}$), for $f_{R0}=-10^{-4}$ and $f_{R0}=-10^{-5}$, the ratio deviates from the $\Lambda$CDM prediction at a level of almost $1\sigma$. For $f_{R0}=-10^{-6}$, due to the screening mechanism, the ratio for massive clusters ($M>10^{13.5}M_{\odot}/h$) is consistent with the $\Lambda$CDM prediction. However, for less massive clusters ($M<10^{13.5}M_{\odot}/h$), the deviations are over $3\sigma$.
In fact, due to the limited box size and number of realizations, the simulations performed in this work tend to over estimate the errors. Our results therefore can be further improved with a larger simulation box and more realizations. The results quoted the above are very conservative.
It is interesting to note that the latest all-sky Planck catalog of Sunyaev-Zeldovich (SZ) sources ~\cite{planck_cluster} has already accumulated over $10^3$ clusters. The clusters are distributed over a large area of the sky. If the pointed X-ray follow-up information is available in the future, the catalog can be used to test $f(R)$ gravity. Further, upcoming the eROSITA survey~\cite{eROSITA} will have the ability to discover over $10^5$ clusters. The shot noise of the measured power spectrum can be reduced significantly.

Our idea can also be further extended. It is interesting to know that the ratio  $P_{\phi}/P_{\phi_N}$ is strongly scale dependent especially when $k<0.2h/{\rm Mpc}$ (see. Fig.~\ref{PPNratio}), on which scales the galaxy bias is expected to be approximately scale independent. Unlike the halo number density field, the total number density field of galaxies, including both central and satellite galaxies, ought to have a close tie with the effective density field. This is because, for massive halos $M>10^{13}M_{\odot}/h$, the total number of satellite galaxies formed inside such halos should relate to the effective mass rather than the true mass of the halos. Thus, if we work with the assumption that galaxies are tracers of the effective density field then their clustering properties can be used to infer the power spectrum of the potential $P_{\phi}$. The power spectrum of the lensing potential $P_{\phi_N}$ can then be inferred from weak lensing data. Future galaxy surveys such as the Euclid mission~\cite{Euclid} will measure both of these statistics in the same region of the sky on large scales. By combining these two pieces of information we may be able to place robust constraints on $f(R)$ gravity. However, this method requires an understanding of galaxy formation in $f(R)$ gravity. We therefore shall elaborate on this idea in detail in future work.

%If we consider galaxies as tracers of the effective density field (or equivalently the potential field $\phi$) from which we can infer $P_{\phi}$, combining weak galaxy lensing surveys from which we can infer $P_{\phi_N}$, we may place robust constraints on $f(R)$ gravity simply using the information in power spectra on a relatively large scale from future galaxy surveys such as the Euclid mission~\cite{Euclid}. We shall elaborate on this idea in detail in future work.

\begin{acknowledgments}
JHH would like to thank Jun Koda for sharing his code and Benjamin R. Granett for helpful discussions. JHH acknowledges support of the Italian Space Agency (ASI), through contract agreement I/023/12/0.
AJH acknowledge support of the European Research
Council through the Darklight ERC Advanced Research Grant (291521).

\end{acknowledgments}

\appendix

\section{Inequalities}
\label{appA}
In this appendix, we provide a strict proof of Eq.~(\ref{strongthinshell}). We shall start with the case of a single particle and then generalize it to scalar fields.

In the linear regime, $\delta R$ can be linearised as
\begin{equation}
\delta R \approx \frac{1}{f_{RR}}\delta f_R=3c^2m_{\rm eff}^2\delta f_R\quad,
\end{equation}
where $m_{\rm eff}^2 = 1/(3c^2f_{RR})$ is the effective mass for the scalar field.
For a point mass particle, under the boundary conditions $\lim_{r\rightarrow+\infty} \delta f_R=\lim_{r\rightarrow+\infty} \phi_N=0$, the linearized Eq.(\ref{frpoisson}) and Eq.(\ref{poissonN}) have solutions
\begin{equation}
\begin{split}
\delta f_R &= \frac{2Gm}{3c^2}\frac{e^{-m_{\rm eff} r}}{r}\quad,\\\label{Greenfunction}
\phi_N&=-\frac{Gm}{r}\quad.
\end{split}
\end{equation}
where $m$ is the mass of the particle. The potential $\delta f_R$ is of the Yukawa type which will be suppressed on scales above the Compton length $\lambda_c = 1/m_{\rm eff}$. From Eq.(\ref{N_phi}), we obtain
\begin{equation}
\phi = -\frac{Gm}{r}-\frac{Gm}{3}\frac{e^{-m_{\rm eff} r}}{r}\label{phi_p}\quad.
\end{equation}
On large scales $r\gg\lambda_c$, therefore, $$e^{-m_{\rm eff} r}\rightarrow 0\quad, $$
the gravity $\phi$ goes back to the standard gravity $\phi\sim \phi_N$. However, on small scales $r\ll\lambda_c$
$$e^{-m_{\rm eff} r}\rightarrow 1\quad, $$ $f(R)$ gravity has a $1/3$ enhancement relative to standard gravity $\phi\sim \frac{4}{3}\phi_N$. $f(R)$ gravity is therefore invalid in the linear regime regardless of the functional form of $f(R)$.

Further, in the large-field limit~\cite{simulation,Fabian}, $m_{\rm eff}\rightarrow 0$, the exponential term in Eq.(\ref{Greenfunction}) becomes $$e^{-m_{\rm eff} r}\rightarrow 1 \quad.$$
$|\delta f_R|$ obtains its maximal value. From Eq.(\ref{phi_p}), in this extreme case, we have
\begin{equation}
c^2\left|\delta f_R\right| \approx \left|-\frac{\phi}{2}\right|\quad.\label{fr_max}
\end{equation}
On the other hand, in the small-field limit \cite{simulation}, $m_{\rm eff}\rightarrow +\infty$ (e.g. the early Universe, when perturbations are small) , the potential $\delta f_R$ will be significantly suppressed
\begin{equation}
c^2\left|\delta f_R\right| \ll \left|-\frac{\phi}{2}\right|\quad.
\end{equation}
In general cases, for the linearized equations, given a finite volume $V$ of the density field, under the boundary condition $$\lim_{|\vec{x}|\rightarrow +\infty}u(\vec{x})=0\quad,$$
the potential is simply the superposition of the potentials generated by local density fields,
\begin{equation}
u(\vec{x})=\iiint\limits_V d\vec{x}'\,G(\vec{x},\vec{x}')\delta \rho(\vec{x}')\quad, \label{superposition}
\end{equation}
where $u(\vec{x})$ standards for the scalar fields $c^2\delta f_R$,$\phi_N$ and $\phi$, respectively. $G(\vec{x},\vec{x}')$ is the Green's function which is given by
\begin{equation}
G(\vec{x},\vec{x}')=
\begin{cases}
\frac{2G}{3|\vec{x}-\vec{x}'|}e^{-m_{\rm eff} |\vec{x}-\vec{x}'|}\quad,&\text{($c^2\delta f_R$),}\\
-\frac{G}{|\vec{x}-\vec{x}'|}\quad,&\text{($\phi_N$),}\\
-\frac{G}{|\vec{x}-\vec{x}'|}-\frac{G}{3|\vec{x}-\vec{x}'|}e^{-m_{\rm eff} |\vec{x}-\vec{x}'|}\quad,&\text{($\phi$).}
\end{cases}\label{Green}
\end{equation}
In high-density regions $\delta \rho\gg 1$, the contribution from the low-density regions ($\delta \rho\lesssim 0$) to the total scalar field compared to the local contribution from the high-density region itself, can be neglected. From the Green's functions Eq.~(\ref{Green}), it follows that
\begin{equation}
c^2\left|\delta f_R\right| \leq \left|-\frac{\phi}{2}\right|\leq |-\frac{2}{3}\phi_N|\quad.
\end{equation}
The above inequality holds everywhere in high-density regions~\cite{frsim}.
However, in low-density regions, the inequality Eq.(\ref{strongthinshell}) may not hold everywhere because the local contribution from the under-dense regions ($\delta \rho<0$) may cancel out the contribution from distant high-density regions to the local total scalar field. $\delta f_R$ and $-\phi_N$ may have different signs. The absolute values of $\delta f_R$ and $-\phi_N$ may have very complicated relations.

\section{Alias sum}
\label{sect:alias}

Let $u(\vec{x})$ be a continuous scalar field. The Fourier transform and its inverse transform are given by
\begin{equation}
\begin{split}
U(\vec{k})&=\int d^3\vec{x}~u(\vec{x})~e^{-i\vec{k}\cdot \vec{x}}\quad,\\
u(\vec{x})&=\int \frac{d^3\vec{k}}{(2\pi)^3}~U(\vec{k})~e^{i\vec{k}\cdot \vec{x}}\quad.\label{FT}
\end{split}
\end{equation}
The two-point correlation function is defined by
\begin{equation}
\begin{split}
&\left\langle u(\vec{x}_1)u(\vec{x}_2)^*\right\rangle\\
=&\int\frac{d^3\vec{k}_1d^3\vec{k}_2}{(2\pi)^6}\left\langle U(\vec{k}_1)U(\vec{k}_2)^*\right\rangle e^{i(\vec{k}_1\cdot\vec{x}_1-\vec{k}_2\cdot\vec{x}_2)}\quad.\label{real2point}
\end{split}
\end{equation}
If we define the power spectrum of the scalar field $u(\vec{x})$ as
\begin{equation}
\left\langle U(\vec{k}_1)U(\vec{k}_2)^*\right\rangle\equiv(2\pi)^3\delta(\vec{k}_1-\vec{k}_2)P_u(k)\quad,
\end{equation}
then the two point correlation function can be written as
\begin{equation}
\begin{split}
\xi(\vec{x}_1-\vec{x}_2)&\equiv \left\langle u(\vec{x}_1)u(\vec{x}_2)^*\right\rangle\\
&=\frac{1}{(2\pi)^3}\int d^3\vec{k} P_u(k)e^{i\vec{k}\cdot(\vec{x}_1-\vec{x}_2)}\quad.\label{twopoint}
\end{split}
\end{equation}

In practice, we can only treat the continuous scalar field $u(\vec{x})$ on discrete grids $\vec{x}_g=\frac{L}{N_g}\hat{n}\, ,$
where $N_g$ is the number of grid cells in one dimension, $L$ is the box size, and $\hat{n}=n_x\hat{x}+n_y\hat{y}+n_z\hat{z}$ describes the position of grid points with $n_x, n_y, n_z$ being integers.
On the discrete grids, the integration of Eq.~(\ref{FT}) can be approximated as a sum over cells with volume $dx^3\approx \frac{L^3}{N_g^3}$
\begin{equation}
\tilde{U}(\vec{k})=\frac{L^3}{N_g^3}\sum_{\hat{n}'}u_{\vec{x}_g}e^{-i\vec{k}\cdot \vec{x}_g}\quad,\label{DFT}
\end{equation}
where $u_{\vec{x}_g}=u(\vec{x})|_{\vec{x}=\vec{x}_g}$. The inverse Fourier transform of Eq.~(\ref{DFT}) can be presented as
\begin{equation}
u_{\vec{x}_g}=\int_0^{2k_N}\int_0^{2k_N}\int_0^{2k_N}\frac{dk_xdk_ydk_z}{(2\pi)^3} \tilde{U}(\vec{k})e^{i\vec{k}\cdot \vec{x}_g}\quad,\label{IDFT}
\end{equation}
where $k_N=\frac{\pi N_g}{L}$ is the Nyquist frequency.
The relation between $\tilde{U}(\vec{k})$ and $U(\vec{k})$ can be obtained by noting that
\begin{equation}
\begin{split}
&u_{\vec{x}_g}=u(\vec{x})|_{\vec{x}=\vec{x}_g}=\int \frac{dk^3}{(2\pi)^3} U(\vec{k})e^{i\vec{k}\cdot \vec{x}_g}\\
=&\int_0^{2k_N}\int_0^{2k_N}\int_0^{2k_N}\frac{dk_xdk_ydk_z}{(2\pi)^3}\sum_{\hat{n}} U(\vec{k}+2k_N\hat{n})e^{i\vec{k}\cdot \vec{x}_g}\quad,\label{alias}
\end{split}
\end{equation}
where in the last equality we have used
\begin{equation}
e^{i\vec{k}\cdot\vec{x}_g}=e^{i(\vec{k}+2k_N\hat{n})\cdot\vec{x}_g}\quad.
\end{equation}
Comparing Eq.~(\ref{alias}) with Eq.~(\ref{IDFT}), we obtain
\begin{equation}
\tilde{U}(\vec{k})=\sum_{\hat{n}} U(\vec{k}+2k_N\hat{n})\quad.\label{PKalias}
\end{equation}
The discrete Fourier transform is simply the sum of replicas of the continuous Fourier transform. This result is known as the ``alias sum''.

Next, we evaluate the two point correlation Eq.~(\ref{real2point}) on discrete grids.
Given that $dk_1^3=dk_2^3\approx(\frac{2\pi}{L})^3$, Eq.~(\ref{twopoint}) can be approximated as
\begin{equation}
\begin{split}
&\left\langle u(\vec{x}_1)u(\vec{x}_2)^*\right\rangle\\
\approx&\frac{1}{L^6}\sum_{\hat{k}_1,\hat{k}_2}\left\langle U_{\hat{k}_1}U_{\hat{k}_2}^*\right\rangle e^{i(\hat{k}_1\cdot \hat{x}_1-\hat{k}_2\cdot{x}_2)}\\
=&\frac{1}{(2\pi)^3}\sum_{\hat{k}}\left(\frac{2\pi}{L}\right)^3\frac{\left\langle |U_{\hat{k}}|^2\right\rangle}{L^3}e^{i\hat{k}\cdot (\vec{x}_1-\vec{x}_2)}\\
\approx&\frac{1}{(2\pi)^3}\int dk^3 \frac{\left\langle |U_{\vec{k}}|^2\right\rangle}{L^3}e^{i\hat{k}\cdot (\vec{x}_1-\vec{x}_2)}
\quad,\label{DPK}
\end{split}
\end{equation}
where $\hat{k}=\frac{2\pi}{L}\hat{n}$ indicates the discrete grids in the Fourier space and $U_{\hat{k}}=U(\vec{k})|_{\vec{k}=\hat{k}}$. In the above derivations, we have used
\begin{equation}
\left\langle U_{\hat{k}_1}U_{\hat{k}_2}^*\right\rangle\equiv\left\langle |U_{\hat{k}}|^2\right\rangle\delta^{D}_{\hat{k}_1,\hat{k}_2}\quad,
\end{equation}
so that the correlation function is only dependent on the spacial separation $\vec{r}=\vec{x}_2-\vec{x}_1$.

Comparing Eq.~(\ref{DPK}) with Eq.~(\ref{twopoint}), we have
\begin{equation}
P_u(\hat{k})=\frac{\left\langle |U_{\hat{k}}|^2\right\rangle}{L^3}\quad.
\end{equation}

Using Eq.~(\ref{PKalias}), we have
\begin{equation}
\begin{split}
P^{FFT}_u(\hat{k})&\equiv\frac{\left\langle |\tilde{U}_{\hat{k}}|^2\right\rangle}{L^3}\\
&=\frac{L^3}{N_g^6}\left\langle {\rm FFT}[u_{\vec{x}_g}]^2\right\rangle\\
&=\frac{\sum_{\hat{n}}\left\langle |U_{\hat{k}+2k_N\hat{n}}|^2\right\rangle}{L^3}\\
&=\sum_{\hat{n}}P_u(\hat{k}+2k_N\hat{n})\quad,
\end{split}
\end{equation}
where ${\rm FFT}[u_{\vec{x}_g}]=\sum_{\hat{n}}u_{\vec{x}_g}e^{-i\hat{k}\cdot \vec{x}_g}$ is the fast Fourier transform of the discrete grid points $u_{\vec{x}_g}$.

Under the assumption of ergodicity, the ensemble average can be replaced by a spatial average. The isotropic power spectrum can be estimated by
\begin{equation}
P^{FFT}_u(k)=\frac{1}{L^3 N_k}\sum_{k\in \Delta k} |\tilde{U}_{k}|^2\quad,
\end{equation}
where  $N_k$ is the number of modes which fall into the spherical shell $\Delta k$ at $k$.

\end{document}